\documentclass[amsmath,amssymb,prc,final,showpacs,twocolumn]{revtex4-1}
 
\usepackage{bm,hyphenat,xspace}
\usepackage{graphicx,epsfig}

\newcommand {\mcu}{\mathcal{U}}

\begin{document}

\title {Breakup of a nucleus with two weakly bound neutrons on a proton target: 
Single-scattering approximation}
 
\author{A.~Deltuva} 
\affiliation{Centro de F\'{\i}sica Nuclear da Universidade de Lisboa, 
P-1649-003 Lisboa, Portugal }


\received{January 31, 2013}

\begin{abstract}
Three- and four-cluster breakup reactions in the 
${}^{16}$C-proton scattering are studied using three-body
core-neutron-neutron model for ${}^{16}$C.
Single-scattering approximation (SSA) of 
 four-particle equations for transition operators 
is used to calculate three- and four-cluster breakup amplitudes
 at 200 and 300 MeV/nucleon energy
near proton-neutron ($pn$) quasi free scattering (QFS) conditions.
The differential cross section is sharply peaked
at $pn$ QFS point and decreases rapidly 
whenever kinematical conditions deviate from $pn$ QFS.
The accuracy of the SSA for the three-cluster breakup 
is estimated from a three-body model and is found
to be as good as 6\% at higher reaction energies in suitable
angular regimes. 
Furthermore, under an additional approximation the three-cluster
breakup amplitude factorizes into the $pn$ transition operator and the
overlap integral of two- and three-particle bound states. That
approximation usually reduces the cross section, in some cases even up
to 10\%.
\end{abstract}

 \maketitle

\section{Introduction \label{sec:intro}}

The structure and properties of exotic short-living nuclei are usually
studied through their reactions with stable targets. 
On the theoretical side,
the reliability of the extracted information depends on the
validity of the assumed interaction model and on the accuracy of the
theoretical methods used to solve the respective few- or many-body
problem. Among the standard tools for the analysis of 
breakup reactions are the distorted wave Born approximation,
the eikonal approximation \cite{baye:09a},
time-dependent models  \cite{esbensen:95a},
 and the more sophisticated continuum-discretized coupled-channels 
(CDCC) method \cite{austern:87}.
For reactions involving effective two-cluster systems like deuteron or
one-neutron halo nuclei that
are treated as three-body problems (two-cluster nucleus plus a target)
also the exact Faddeev-type formalism \cite{faddeev:60a,alt:67a}
in the momentum-space framework has successfully been applied.
 Within this approach, once numerically 
well-converged results are obtained, 
all discrepancies with the experimental data can be 
attributed to the shortcomings of the used  three-body model.
However, practical applications of the Faddeev-type formalism
are limited so far to light projectiles and targets.

On the other hand,  Faddeev-type calculations can be used to check the accuracy 
of the traditional approximate nuclear reaction methods. Indeed,
the comparison \cite{deltuva:07d} of Faddeev-type and CDCC results
revealed that CDCC is a reliable method for
deuteron-nucleus elastic scattering and breakup but  fails 
for the breakup of a one-neutron halo nucleus,
assumed to be a bound state of an inert nuclear core ($A$) 
and a neutron ($n$), on a proton ($p$) target.
More specifically, the peaks of the differential cross section  
in energy and angular distributions of the core $A$
are underpredicted by CDCC. 
 Those peaks are due to $pn$ quasi free scattering (QFS), i.e., 
only the neutron interacts with the proton and is knocked out
from the $(An)$ nucleus whereas the momentum and energy transfers
to the core $A$ vanish and, as a consequence, the core
keeps the velocity of the incoming beam.  Thus, near the $pn$ QFS 
the reaction dynamics is dominated by the $pn$ interaction
which is very difficult to describe well when the three-particle
CDCC wave function is
expanded using the $nA$ bound and selected continuum states
as done in standard CDCC  (the transfer-to-continuum CDCC \cite{moro:06a}
is more successful, but so far it is only available for three-body systems).

For the same reason one may expect the failure of CDCC for the 
breakup of a core and two-neutron nucleus $(Ann)$ on a proton target
near $pn$ QFS kinematics, especially at higher energies relevant for 
the analysis of present day and future experiments.
Although four-body CDCC \cite{matsumoto:06a,gallardo:09a}
and eikonal approximation \cite{baye:09a}
calculations for breakup of two-neutron halo nuclei exist,
to the best of our knowledge, they were not applied in the 
$pn$ QFS regime.
Furthermore, additional technical difficulties arise in CDCC
if two-body bound states $(An)$ exist \cite{moro:pc}. On the other
hand, the rigorous treatment of the four-particle scattering problem within the
Faddeev-type framework was successfully performed only for the
four-nucleon system so far \cite{deltuva:12c,lazauskas:12a}. The extension 
of the Faddeev-type method to distinguishable particles and high energies
is very difficult technical challenge, although
 in principle it may be feasible. On the other hand, such a huge effort
may be unnecessary for the description of particular breakup reactions 
where the CDCC
is expected to fail, i.e., for $p+(Ann)$ breakup near $pn$ QFS kinematics.
The studies of ${}^{11}$Be breakup on a proton target at
200 MeV/nucleon reaction energy \cite{crespo:08a} around the
$pn$ QFS kinematics revealed that the single scattering
approximation (SSA) reproduces quite accurately the results
of the full Faddeev-type calculations for not too small neutron scattering
angles. Thus, one may expect the SSA to be reasonable also
in the four-particle system at similar kinematical conditions.
We therefore aim  at developing the four-particle
SSA to study the breakup reactions $p+(Ann) \to p + n + (An)$ and
$p+(Ann) \to p + n + n + A$ at higher energies.
The numerical results use $p+{}^{16}$C reactions as example.
Unfortunately, no experimental data is available.
 In the present work we will concentrate
 on theoretical aspects of the
three-cluster breakup where one may create 
effective three-body model $p+(Bn) \to p + n + B$ with $B=(An)$.
Physically such a model is inadequate for weakly bound $(An)$,
but it allows to perform full three-body Faddeev-type calculations
and, by comparing with the respective SSA, identify the kinematical
regimes where the SSA is reasonable, and estimate 
its accuracy.

In Sec.~\ref{sec:eq} we derive the SSA for the breakup amplitudes.
In Sec.~\ref{sec:v} we describe the employed interaction model.
We study the  three-cluster breakup in Sec.~\ref{sec:3c} 
and four-cluster breakup in Sec.~\ref{sec:4c}.
Summary is given  in Sec.~\ref{sec:sum}.

\section{Breakup amplitudes \label{sec:eq}}

We consider the four-particle system
interacting via short-range pairwise potentials $v_j$ 
where $j$ labels the respective pair.
We do not calculate the four-particle wave function explicitly
but work in the momentum-space and 
use the integral form of the scattering equations as proposed
by Alt, Grassberger and Sandhas  (AGS) \cite{grassberger:67}.
The AGS equations are equivalent to 
the Faddeev-Yakubovsky equations \cite{yakubovsky:67}
but are formulated
for the  four-particle transition operators $\mcu_{\sigma \rho}^{ji}$, i.e.,
\begin{equation} \label{eq:AGSgen}
\mcu_{\sigma \rho}^{ji} = 
(G_0\, t_i\, G_0)^{-1}\,
\bar{\delta}_{\sigma \rho} \, \delta_{ji} 
  + \sum_{\gamma k} \bar{\delta}_{\sigma \gamma}  U_{\gamma}^{jk} G_0\, t_k\, 
G_0 \, \mcu_{\gamma \rho}^{ki},
\end{equation}
The components of the operators are distinguished
by the two-cluster partition and by the three-cluster partition;
18 different combinations are possible. 
The two-cluster partitions,
denoted by Greek letters, are either of 1+3 or 2+2 type.
The  three-cluster partitions, denoted by Latin letters,
 are of 2+1+1 type and therefore are fully
characterized by the pair  of particles in the composite cluster.
Obviously,  the pairs $i$, $j$ and $k$ must be
 internal to the respective two-cluster partitions, i.e.,
$i \subset \rho$, \, $j,k \subset \gamma$ and $j \subset \sigma$,
whereas $\bar{\delta}_{\sigma \rho} = 1 - {\delta}_{\sigma \rho}$. 
The components $\mcu_{\sigma \rho}^{ji}$ with all allowed $j$ and $i$
describe the transition
from the initial  two-cluster partition $\rho$ to the 
final  two-cluster partition $\sigma$ \cite{deltuva:07a}.
In our example of the $p+(Ann)$ scattering the initial
 two-cluster partition $\rho$ is $p+(Ann)$ with three internal pairs, i.e.,
$(nn)$ and twice $(An)$. Other two-cluster partitions of 
the 1+3 type are $A+(nnp)$ and twice $n+(npA)$
while those of the 2+2 type are $(pA)+(nn)$ and twice $(pn)+(An)$. 
 Note that the components of the transition operators
exist even if the particles in the corresponding cluster
do not bind as in the case of $(pA)+(nn)$; 
 those components contribute to breakup reactions. 

The  free resolvent at the available system energy $E$ is given by
\begin{equation} \label{eq:G0}
G_0 = (E+i0-H_0)^{-1}
\end{equation}
where $H_0$ is the kinetic energy operator.
The two-particle transition operator 
\begin{equation} \label{eq:t2b}
t_j = v_j + v_j G_0 t_j
\end{equation}
sums up the interactions for pair $j$ up to all orders.
Furthermore, all interactions within each two-cluster subsystem $\gamma$
lead to the respective subsystem transition operators
\begin{gather}
U_{\gamma}^{jk} = G^{-1}_0 \, \bar{\delta}_{jk} + \sum_i \bar{\delta}_{ji}
\, t_i \, G_0 \, U_{\gamma}^{ik}.
\end{gather}

As derived in Ref.~\cite{deltuva:12e}, the amplitude for the 
three-cluster breakup of the initial two-cluster state is
\begin{equation} \label{eq:AGS3}
 \langle \Phi_{j} |  T_{j \rho} | \Phi_{\rho} \rangle
=  \sum_{\gamma ki} \langle \Phi_j |
U_{\gamma}^{jk} G_0\, t_k\, G_0 \, 
\mcu_{\gamma \rho}^{ki} | \phi_{\rho}^i \rangle.
\end{equation}
The energy parameter in the operators of Eq.~(\ref{eq:AGS3})
is $E =  \epsilon_{\rho} + p_{\rho}^2/2\mu_\rho $ 
with  $\epsilon_{\rho}$ being the energy of the bound state
in the partition $\rho$ and $\mu_{\rho}$ being the 
respective two-cluster reduced mass.
The initial asymptotic channel state $| \Phi_{\rho} \rangle$
is a product of the bound state  wave function 
in the partition $\rho$  and the plane 
wave with momentum  $\mathbf{p}_{\rho}$ between the two clusters.
$| \Phi_{\rho} \rangle =  \sum_{i} | \phi_{\rho}^i \rangle$
is decomposed into its Faddeev components satisfying
\begin{equation} \label{eq:phi}
| \phi_{\rho}^i \rangle = G_0 \sum_{j} \bar{\delta}_{ij} t_j 
| \phi_{\rho}^j \rangle
\end{equation}
and normalized such that
$\langle \Phi_{\sigma} | \Phi_{\rho} \rangle = \delta_{\sigma \rho} \,
\, \delta(\mathbf{p}_{\sigma}-\mathbf{p}_{\rho})$.
The asymptotic three-cluster 
state $|\Phi_{j} \rangle $ is an eigenstate of the
channel Hamiltonian $H_0 + v_j$ with the eigenvalue $E$. 
It is given by the bound state
wave function for the pair $j$ times two plane waves 
corresponding to the relative motion of three free clusters.

The amplitude for the four-cluster breakup is taken over 
from  Ref.~\cite{deltuva:12a}, i.e.,
\begin{equation} \label{eq:AGS0}
 \langle \Phi_{0} |  T_{0 \rho} | \Phi_{\rho} \rangle
=  \sum_{\gamma jki} \langle \Phi_{0} |
t_j \, G_0 \,  U_{\gamma}^{jk} G_0\, t_k\, G_0 \, 
\mcu_{\gamma \rho}^{ki} | \phi_{\rho}^i \rangle.
\end{equation}
The four-cluster channel state $| \Phi_{0} \rangle$
is an eigenstate of $H_0$ with eigenvalue $E$;
it is a product of three plane waves
(each is normalized to the Dirac $\delta$-function)
corresponding to the relative motion of four free particles.

In the SSA, i.e., keeping only the terms of the first order in
two-particle transition operators (\ref{eq:t2b}), the
three- and four-cluster breakup amplitudes become
\begin{subequations}  \label{eq:ssa}   
\begin{align}  \label{eq:ssa3}
\langle \Phi_{j} |  T_{j \rho}^{\rm SSA} | \Phi_{\rho} \rangle
=  \sum_{k}  \bar{\delta}_{k\rho} 
\langle \Phi_j | t_k | \Phi_{\rho} \rangle, \\
\label{eq:ssa4}
\langle \Phi_{0} |  T_{0 \rho}^{\rm SSA} | \Phi_{\rho} \rangle
=  \sum_{k}  \bar{\delta}_{k\rho} 
\langle \Phi_0 | t_k | \Phi_{\rho} \rangle.
\end{align}
\end{subequations}
Here $\bar{\delta}_{k\rho} $ is  0  if $k \subset \rho$ 
and 1 otherwise.
Thus, in Eqs.~(\ref{eq:ssa}) the summation runs over 
all pairs that are external to the initial state partition $\rho$. Note that
 Eq.~(\ref{eq:ssa3})  is the amplitude for the direct three-cluster breakup
and not for rearrangement breakup, i.e., the final bound pair $j$
is internal to the initial partition $\rho$.

In the following we consider the initial partition $\rho$ to be of
the 1+3 type, i.e., 1(234) with particle 1 as spectator and
the bound state of particles (234). In such case the amplitudes (\ref{eq:ssa})
have three contributions 
$\sum_{k}  \bar{\delta}_{k\rho} t_k = t_{12} + t_{13} + t_{14}$
corresponding to the interactions of the spectator particle 1
with each particle in the cluster (234) but no interactions within the cluster.

We start with the four-cluster breakup amplitude (\ref{eq:ssa4}).
Let $m_a$ be the mass of the particle $a$ and $\mathbf{k}_a$
its final-state momentum with $a=1,2,3,4$. 
Furthermore,  the initial momentum of particle 1 and of cluster (234)
we denote by $\mathbf{k}_1^i$ and  $\mathbf{k}_\rho^i$, respectively.
Obviously,  $\mathbf{k}_1^i$, $\mathbf{k}_\rho^i$ and $\mathbf{k}_a$
 are related  by  momentum and energy conservation, i.e.,
$\sum \mathbf{k}_a = \mathbf{k}_1^i + \mathbf{k}_\rho^i = \mathbf{K}$
and
$\sum k_a^2/2m_a = \epsilon_{\rho} + {k_1^i}^2/2m_1 + {k_\rho^i}^2/2m_\rho = E$ 
with $m_\rho = m_2+m_3+m_4$.
We do not assume a particular frame, thus, the results are valid
with the energy $E$ and momenta given in any frame.
The explicit momentum-dependence of the $ t_{12}$ term is
\begin{equation} \label{eq:t4}
 \langle \Phi_{0} |  t_{12} | \Phi_{\rho} \rangle
=  \langle \mathbf{p}_{12} | t_{12}(e_{12}+i0)| \mathbf{p}_{12}' \rangle
\langle \mathbf{q}_2 \mathbf{p}_{34} |\Phi_{\rho} \rangle
\end{equation}
with 
\begin{subequations}  \label{eq:p}   
\begin{align}  \label{eq:p1}
\mathbf{p}_{ab} = {} & \frac{m_b\mathbf{k}_a-m_a\mathbf{k}_b}{m_a+m_b}, \\
\mathbf{p}_{1b}' = {} & \frac{m_b\mathbf{k}_1^i-m_1\mathbf{k}_b'}{m_1+m_b}, \\
\mathbf{k}_b' = {} & \mathbf{k}_1+\mathbf{k}_b-\mathbf{k}_1^i ,  \\
\mathbf{q}_2 =  {} & \frac{1}{m_\rho}
[(m_3+m_4)\mathbf{k}_2'-m_2(\mathbf{k}_3+\mathbf{k}_4)].
\end{align}
\end{subequations}
Alternatively, another set of Jacobi momenta for the (234) subsystem
could be chosen to represent $|\Phi_{\rho} \rangle$.
Here and in the following the wave functions with notationally indicated
momentum dependence refer to the internal motion of the respective
bound cluster; the part corresponding to the free motion between
clusters is taken out. 
The pair transition operator $t_{12}$ depends on 
the energy available for the relative motion of particles 1 and 2, i.e.,
\begin{equation} \label{eq:e4}
e_{12} = E - \frac{k_3^2}{2m_3} - \frac{k_4^2}{2m_4} - 
\frac{(\mathbf{k}_1+\mathbf{k}_2)^2}{2(m_1+m_2)}.
\end{equation}
Due to momentum and energy conservation  $e_{12}  = p_{12}^2/2\mu_{12}$ 
with the reduced mass $\mu_{ab} = m_am_b/(m_a+m_b)$; thus,
the pair transition operator $ t_{12}$ has to be evaluated half-shell.
Furthermore, although for brevity not explicitly indicated in our notation,
the  breakup amplitudes (\ref{eq:ssa}) and pair transition matrices
(\ref{eq:t2b}) are operators in the spin-space
implying summations over all intermediate spin states.
In Eq.~(\ref{eq:t4}) this summation involves only the
initial spin projection of the particle 2.

The momentum-dependence of the $ t_{13}$  and $ t_{14}$ terms
has the same structure and can be easily obtained from
Eq.~(\ref{eq:t4}) by the respective permutation of particle labels.

\begin{figure}[!]
\begin{center}
\includegraphics[scale=0.62]{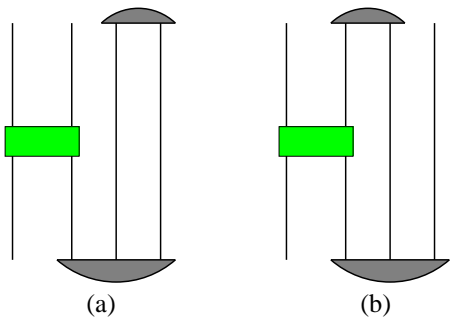}
\end{center} 
\caption{ \label{fig:ss3} (Color online) 
Two types of contributions to the three-cluster
breakup amplitude in the SSA. The two-particle transition operator
is represented by a box while two- and three-particle bound states
are represented by filled arcs. The diagrams (a) and (b) 
correspond to Eqs.~(\ref{eq:t3}) and (\ref{eq:t33}), respectively.}
\end{figure}

Considering the three-cluster breakup we assume that
particles 3 and 4 build the final bound pair with mass $m_B = m_3+m_4$,
total momentum $\mathbf{k}_B$, internal
wave function  $ \langle \mathbf{p}_{34} | \Phi_B \rangle$ 
and energy $\epsilon_{B}$.
The kinematics of the QFS condition for the pair of particles 1 and 2 reads 
$\mathbf{k}_B \approx \mathbf{k}_B^{\rm QFS} = (m_B/m_\rho)\mathbf{k}_\rho^i$.
The most important contribution of the breakup amplitude is the $ t_{12}$ term
represented diagrammatically in Fig.~\ref{fig:ss3} (a). 
It includes the interaction between the spectator and the struck particle,
\begin{equation} \label{eq:t3}
\begin{split}
 \langle \Phi_{B} |  t_{12} | \Phi_{\rho} \rangle
=  & {} \int d^3p_{34} [
\langle \mathbf{p}_{12} | t_{12}(e_{12}+i0)| \mathbf{p}_{12}' \rangle \\
& \times \langle \Phi_B | \mathbf{p}_{34} \rangle
\langle \mathbf{q}_2 \mathbf{p}_{34} |\Phi_{\rho} \rangle].
\end{split}
\end{equation}
The definitions of the involved momenta 
as given in Eqs.~(\ref{eq:p}) are valid also here but
$\mathbf{k}_3 = m_3\mathbf{k}_B/(m_3+m_4) + \mathbf{p}_{34}$ and 
$\mathbf{k}_4 = m_4\mathbf{k}_B/(m_3+m_4) - \mathbf{p}_{34}$
are not independent variables anymore.   The momentum and 
energy conservation changes to
$\mathbf{k}_1 + \mathbf{k}_2 +\mathbf{k}_B  
= \mathbf{k}_1^i + \mathbf{k}_\rho^i = \mathbf{K}$ and 
$ \epsilon_{B} + k_1^2/2m_1 +  k_2^2/2m_2 +  k_B^2/2m_B 
= \epsilon_{\rho} + {k_1^i}^2/2m_1 + {k_\rho^i}^2/2m_\rho = E$.
The energy $e_{12}$ as defined in Eq.~(\ref{eq:e4}) is valid as well
but it is more appropriate to express it via $\mathbf{p}_{ab}$, i.e.,
\begin{equation}
\label{eq:e3}
e_{12} = \frac{p_{12}^2}{2\mu_{12}} + \epsilon_{B} - \frac{p_{34}^2}{2\mu_{34}}. 
\end{equation}
Thus, this time the two-particle transition operator
has to be evaluated fully off-shell. 
More importantly, Eq.~(\ref{eq:e3}) demonstrates that
$ t_{12}(e_{12}+i0)$
 depends on the integration variable $p_{34}$ and therefore
cannot be taken out of the integral in Eq.~(\ref{eq:t3}).
An additional approximation in the energy-dependence of $ t_{12}$
is needed to factorize the
three-cluster breakup amplitude (\ref{eq:t3}) into $ t_{12}$ and
the overlap integral (OI)
\begin{equation} \label{eq:oi}
\chi_\rho^B(\mathbf{q}_2) = 
\int d^3p_{34} \langle \Phi_B | \mathbf{p}_{34} \rangle
\langle \mathbf{q}_2 \mathbf{p}_{34} |\Phi_{\rho} \rangle.
\end{equation}
We introduce the overlap integral approximation (OIA) of the
three-cluster breakup amplitude as
\begin{equation} \label{eq:t3oi}
 \langle \Phi_{B} |  T_{B \rho}^{\rm OIA}  | \Phi_{\rho} \rangle
= \langle \mathbf{p}_{12} | \, t_{12}\left(\frac{p_{12}^2}{2\mu_{12}}+i0 \right)
| \mathbf{p}_{12}' \rangle \,  \chi_\rho^B(\mathbf{q}_2).
\end{equation}
Under this particular approximation the two-particle transition operator
needs to be evaluated half-shell only and the  four-particle SSA becomes 
formally a three-particle SSA since the amplitude (\ref{eq:t3oi})
has exactly the structure of the SSA breakup amplitude for the three-particle
system (1+2+B) where the (2B) bound state wave function is
replaced by the OI $\chi_\rho^B(\mathbf{q}_2)$.

The $ t_{13}$ and $ t_{14}$ terms of the three-cluster
breakup amplitude (\ref{eq:ssa3}) have the structure 
diagrammatically represented  in Fig.~\ref{fig:ss3} (b).
They include the interactions
between the spectator and the particles that remain bound in the pair B, i.e.,
\begin{equation} \label{eq:t33}
\begin{split}
 \langle \Phi_{B} |  t_{13} | \Phi_{\rho} \rangle
=  & {} \int d^3p_{34} [
\langle \mathbf{p}_{13} | t_{13}(e_{13}+i0)| \mathbf{p}_{13}' \rangle \\
& \times \langle \Phi_B | \mathbf{p}_{34} \rangle
\langle \tilde{\mathbf{q}}_2 \tilde{\mathbf{p}}_{34} |\Phi_{\rho} \rangle]
\end{split}
\end{equation}
with $\tilde{\mathbf{q}}_2 =
[(m_3+m_4)\mathbf{k}_2-m_2(\mathbf{k}_3'+\mathbf{k}_4)]/m_\rho$, 
 $\tilde{\mathbf{p}}_{34}= (m_4\mathbf{k}_3'-m_3\mathbf{k}_4)/(m_3+m_4)$,
and $e_{13}$ defined according to Eq.~(\ref{eq:e4}) 
with the permutation  $(2 \leftrightarrow 3)$.
Thus, all relative
momenta in Eq.~(\ref{eq:t33}) depend on the integration variable 
$\mathbf{p}_{34}$ and therefore no simplifications are possible.
However, our numerical calculations revealed that in the regime
relevant for SSA, i.e., near $pn$ QFS, the contributions of
the type (\ref{eq:t33}) are very small as compared to (\ref{eq:t3})
and can be safely neglected. There are good physics reasons for this
since, as mentioned,  $ t_{13}$ and $ t_{14}$ terms describe the spectator
interaction with the particles that form the
bound pair B but not with the struck particle. 
The suppression of contribution (\ref{eq:t33}) comes from 
high momentum components of the involved wave functions
while in Eq.~(\ref{eq:t3}) low-momentum components are decisive.

For $p+(Ann)$ breakup  reactions considered in this work
two additional remarks regarding the
symmetrization and the long-range Coulomb interaction are needed.

First, the channel states in Eqs.~(\ref{eq:ssa}) have to be
antisymmetric under exchange of the two neutrons.
For the initial state $| \Phi_{\rho} \rangle$
this is achieved by including only antisymmetric two-neutron partial waves
and keeping only two independent Faddeev components $| \phi_{\rho}^i \rangle$
when solving the three-particle bound state problem.
The antisymmetrized final asymptotic three- and four-cluster states 
to be used in Eqs.~(\ref{eq:ssa}) are
\begin{equation} \label{eq:asym}
 | \Phi_{x}^s \rangle
= \frac{1}{\sqrt{2}}(1-P_{nn})| \Phi_{x} \rangle
\end{equation}
with $P_{nn}$ being the two-neutron permutation operator.
Since  $| \Phi_{\rho} \rangle$ is already antisymmetrized and 
$t_{12} + t_{13} + t_{14}$ is symmetric under exchange of the neutrons,
the antisymmetrization (\ref{eq:asym}) simply yields an additional
$\sqrt{2}$ factor for the amplitudes (\ref{eq:ssa}).

Second, the AGS equations for the transition operators are defined
only with short-range potentials $v_j$. The Coulomb
interaction (with no more than two charged clusters)
can be included using the method of screening and
renormalization \cite{alt:80a,deltuva:05d,deltuva:07b}
but only in the full form of AGS equations where the
unscreened limit for the renormalized amplitudes exists.
This is not the case in the SSA: $t_{pn}$ being
Coulomb-free needs no renormalization while $t_{pA}$ includes Coulomb
and needs renormalization half-shell \cite{alt:80a}.
Thus, strictly speaking, their sum has no unscreened
limit neither with nor without renormalization.
Practically, this is not a problem since, according to our calculations,
 near the QFS conditions where the SSA is expected to be reasonable, the 
contribution of $t_{pA}$ is very small and can safely be neglected.

\section{Interactions  \label{sec:v}}

We take $p+{}^{16}$C$ \to p+n+{}^{15}$C and $p+{}^{16}$C$\to p+n+n+{}^{14}$C 
reactions in inverse kinematics as a working example in the numerical 
calculations of this paper. The ground state of ${}^{14}$C, the core $A$,
is well separated from the excited states (6.093 MeV) while
${}^{15}$C is weakly bound ($\epsilon_{B}= -1.218$ MeV) one-neutron halo nucleus
for which a simple two-body model of core $A$ and neutron $n$
is assumed to be adequate.
The neutron separation energy of ${}^{16}$C is 4.250 MeV
such that core and two-neutron  model of ${}^{16}$C ground state $0^+$
 with $\epsilon_{\rho} = -5.468$ MeV appears to be quite reasonable.

As the $pn$ and $nn$ interactions we take the
charge-dependent Bonn (CD Bonn) potential \cite{machleidt:01a}.
The $nA$ interaction is taken over from Ref.~\cite{deltuva:09d};
it has central and spin-orbit parts of Woods-Saxon shape adjusted
to ${}^{15}$C bound states $2s_{1/2}$ and  $1d_{5/2}$ and
to ${}^{14}$C neutron separation energy in  $1p_{1/2}$.
The deeply-bound Pauli forbidden states $1s_{1/2}$, $1p_{1/2}$, and
$1p_{3/2}$ are projected out as described in Ref.~\cite{deltuva:09d}.
The optical potential as parametrized by 
Koning \& Delaroche \cite{koning} plus screened Coulomb
is used for $pA$ interaction; one could probably find a better
parametrization but, as already mentioned, this is irrelevant since 
$t_{pA}$ yields negligible contribution to SSA near $pn$ QFS.

The above $nn$ and  $nA$ potentials alone do not yield the desired
value  $\epsilon_{\rho} = -5.468$ MeV for the ${}^{16}$C ground state.
A simple way to remedy this shortcoming is to add a  three-body force
(3BF) acting in the $nnA$ subsystem only. In this way it does not affect
the functional form of the SSA breakup amplitudes (\ref{eq:ssa}).
The coordinate-space calculations usually take a central 3BF depending
on hyperradius. Our calculations are in momentum space so we
take a central 3BF depending on hypermomentum $\mathcal{K}$
that in the three-particle (234) subsystem has the form
\begin{equation} \label{eq:3bf}
\langle  \mathbf{k}_2 \mathbf{k}_3 \mathbf{k}_4
| W | \mathbf{k}'_2 \mathbf{k}'_3 \mathbf{k}'_4  \rangle = 
(4\pi)^2 w_3 g(\mathcal{K}^2) g(\mathcal{{K}'}^2)
\end{equation}
with $\mathcal{K}^2 = m_N[k_2^2/m_2+ k_3^2/m_3+k_4^2/m_4
-( \mathbf{k}_2+ \mathbf{k}_3+ \mathbf{k}_4)^2/m_\rho]$ and
$\mathcal{{K}'}^2$ defined analogously where $m_N$ is the average nucleon
mass. Note that $\mathcal{K}^2/2m_N$ is the internal motion kinetic energy 
of the  (234) subsystem. We chose the form factor
$g(\mathcal{K}^2) = \exp{(-\mathcal{K}^2/2\Lambda^2)}$ as Gaussian.
The form (\ref{eq:3bf}) of the 3BF is inspired by the effective
field theory \cite{hammer:07a}.
Once for each value of the cutoff parameter
$\Lambda$ the strength  $w_3$ is adjusted
to reproduce $\epsilon_{\rho} = -5.468$ MeV, the predictions  
become practically independent of $\Lambda$. For example, changing
$\Lambda$ by a factor of 2,  from $2\, {\rm fm}^{-1}$ to $4\, {\rm fm}^{-1}$,
 yields less than 3\% changes in the breakup cross sections.
 Our standard choice is $\Lambda = 3\, {\rm fm}^{-1}$.

With 3BF included the Faddeev components of the three-particle
bound state obey the equation
\begin{equation} \label{eq:phi3}
| \phi_{\rho}^i \rangle = G_0 \sum_{j} \bar{\delta}_{ij} t_j 
| \phi_{\rho}^j \rangle
+ G_0(1+t_iG_0) \eta_{i} W \sum_{j} | \phi_{\rho}^j \rangle
\end{equation}
with $\sum \eta_i =1$. Different $\eta_i$ choices correspond to
different splitting of the 3BF contributions among the Faddeev components
but yield identical $\epsilon_{\rho}$ and $| \Phi_{\rho} \rangle$.

We calculate the two-particle transition operators $t_j$
and the bound state wave functions $| \Phi_{B} \rangle$ and
$| \Phi_{\rho} \rangle$ in the momentum-space partial-wave basis but 
then transform them into the plane-wave representation as needed
in Eqs.~(\ref{eq:t4}-\ref{eq:t33}).
To obtain converged results the two-particle interactions $v_j$ have to be
included up to the two-particle relative orbital angular momentum $L_{\max}^j$.
We find that $L_{\max}^j =11$ for $pn$,  3 for $nn$, and 3 for $nA$
is sufficient.
The test calculations proving the negligible contribution
of $t_{pA}$ used $L_{\max}^j = 20$.

We note that at $e_{12} = \epsilon_{d} $ with $ \epsilon_{d}= -2.223$ MeV 
being the CD Bonn prediction for the
deuteron bound state energy, the $t_{pn}$ transition
operator in the ${}^3S_1 - {}^3D_1 $ partial wave
exhibits the deuteron bound state pole. In the integrals it is treated
by the subtraction technique.

\section{Three-cluster breakup \label{sec:3c}}

We consider the three-cluster breakup where two clusters ($a$ and $b$)
are detected with momenta $\mathbf{k}_a$ and $\mathbf{{k}}_b$
(all single-cluster momenta in this section refer to the lab frame).
The momentum of the undetected cluster $c$ is fully determined
by the momentum conservation. The energy conservation renders
$k_a$ and $k_b$ not independent; for a fixed $k_a$ there may be
up to two solutions for $k_b$, and vice versa.
The five independent kinematic variables for the
fully exclusive fivefold differential cross section are often
chosen as the polar and azimuthal scattering angles 
$\Omega_a = (\Theta_a,\varphi_a)$ and 
$\Omega_b = (\Theta_b,\varphi_b)$ of the two detected particles
and one energy $E_a$, i.e.,
\begin{equation} \label{eq:d5s}
\begin{split}
\frac{d^5\sigma}{d\Omega_a d\Omega_b dE_a} = {} &
 \frac{ (2\pi)^4 m_a m_b m_c k_a k_b^3} 
{V|(m_b+m_c)k_b^2 - m_b(\mathbf{K}-\mathbf{k}_a)\cdot\mathbf{{k}}_b|} \\
& \times \frac{1}{g_i} \sum_{m_s}
| \langle \Phi_{B} |  T_{B \rho} | \Phi_{\rho} \rangle|^2.
\end{split}
\end{equation}
Here $V = |\mathbf{k}_1^i/m_1 - \mathbf{k}_\rho^i/m_\rho|$ is the incoming flux.
The sum runs over  all initial and final spin states,
while $g_i = (2s_1+1)(2s_\rho+1)$ takes care of the  spin averaging
in initial state, $s_1$ ($s_\rho$) being the spin of the 
particle 1 (cluster $\rho$).

In our example of  $p+{}^{16}$C$ \to p+n+{}^{15}$C reaction
we assume that beam of ${}^{16}$C is impinging on target $p$ and
the detected particles are ${}^{15}$C in its ground state ($B$) and $p$.
We compare the differential cross section (\ref{eq:d5s}) calculated
in four different ways, depending on the scattering amplitude
$\langle \Phi_{B} |  T_{B \rho} | \Phi_{\rho} \rangle$. 
The four-particle SSA as given by Eqs.~(\ref{eq:t3}-\ref{eq:e3})
is labeled SSA-4b. 
Its approximation (\ref{eq:t3oi}) is labeled OIA-4b.
The importance of higher order interactions between the three clusters
can be estimated by creating an effective three-body model where
the composite cluster $B$ is treated as a single inert particle.
The $nB$ potential is real in $0^+$ wave and supports bound state 
with energy $\epsilon_{\rho} - \epsilon_{B}$, while
the $pB$ interaction includes optical Koning \& Delaroche \cite{koning} 
and Coulomb potentials; the latter is treated using the method of
screening and renormalization \cite{deltuva:05d,deltuva:07d}.
The results obtained by solving full three-body  Faddeev-type equations,
i.e., formally summing up multiple scattering (MS) series up to infinite
order, are labeled MS-3b. The SSA of this model including
only $t_{pn}$ term is labeled SSA-3b. Although such model makes
physically little sense owing to halo nature of $B$, but the
ratio [(MS-3b)$-$(SSA-3b)]/(MS-3b) should be a reasonable
accuracy estimate of SSA-4b. Note that for comparison the results of the 
three-body model are multiplied by 2 to account for the two neutrons
in the original four-particle model.

We concentrate on the kinematic regime near $pn$ QFS, i.e.,
$\mathbf{k}_B \approx \mathbf{k}_B^{\rm QFS} = (m_B/m_\rho)\mathbf{k}_\rho^i$.
In terms of angles and energy this means $\Theta_B \approx 0$ and 
$E_B \approx E_B^{\rm QFS} = ({k}_B^{\rm QFS})^2/2m_B = (m_B/m_\rho)E_\rho^i$
where $E_\rho^i$ is the  beam energy of ${}^{16}$C.
Note that at these conditions also $\mathbf{q}_2 $ vanishes 
in Eq.~(\ref{eq:t3}).

\begin{figure}[!]
\begin{center}
\includegraphics[scale=0.62]{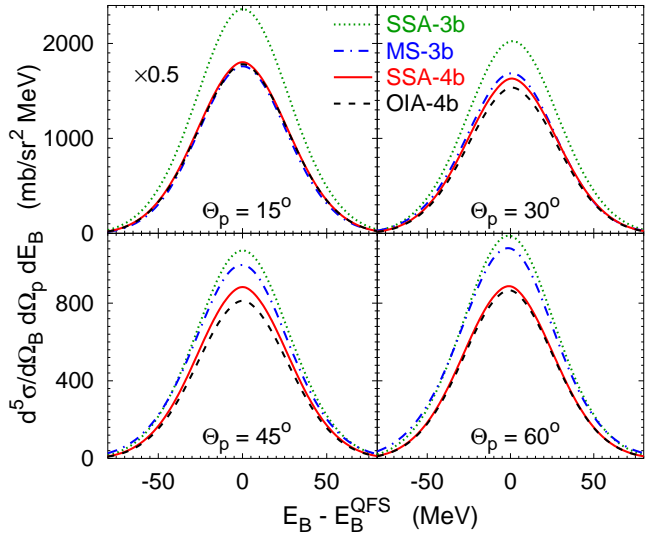}
\end{center} 
\caption{ \label{fig:p300} (Color online) 
Differential cross section for ${}^{16}$C$(p,pn){}^{15}$C reaction in 
inverse kinematics as a function of the final ${}^{15}$C  energy $E_B$.
The energy of the ${}^{16}$C beam is 300 MeV/nucleon. 
${}^{15}$C and proton are detected at angles
$(\Theta_B=0^\circ,\varphi_B=0^\circ)$ and 
$(\Theta_p=15,30,45,60^\circ,\varphi_p=180^\circ)$.
Results of SSA-3b (dotted), MS-3b (dotted-dashed), SSA-4b (solid)
and OIA-4b (dashed) calculations are compared. }
\end{figure}

In Fig.~\ref{fig:p300} we show the results at $E_\rho^i/16 = 300$ MeV.
We fix $\Theta_B = 0^\circ$, $\varphi_B= 0^\circ$, $\varphi_p= 180^\circ$, 
 and vary $\Theta_p$. 
In all used approaches the differential cross section peaks quite sharply at
(or very near to) $E_B = E_B^{\rm QFS}$. This is due to sharp increase
of the $s$-wave components of the
bound state wave functions for vanishing relative momenta.
At $\Theta_p = 15^\circ$ and $30^\circ$ there is significant difference
between MS-3b and SSA-3b indicating that the SSA is not reliable in this
regime. On the contrary, at $\Theta_p = 45^\circ$ and $60^\circ$
the agreement between MS-3b and SSA-3b gets better, of the order of
 6\% at the peak. Thus, for these kinematic configurations 
SSA-4b should be of a comparable accuracy. The SSA-4b results,
taking into account the three-particle structure of ${}^{16}$C, 
are considerably lower than the ones of SSA-3b where this aspect is 
neglected. The OIA-4b, involving an
additional approximation in the energy-dependence of $t_{pn}$,
underestimates the SSA-4b up to 9\%, most sizably at $\Theta_p = 45^\circ$.

\begin{figure}[!]
\begin{center}
\includegraphics[scale=0.62]{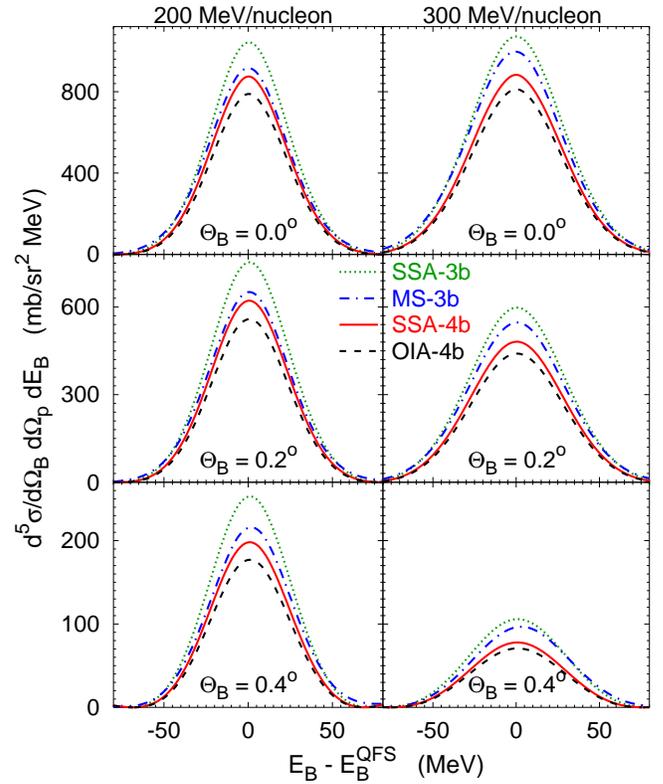}
\end{center} 
\caption{ \label{fig:p2-300} (Color online) 
Same as Fig.~\ref{fig:p300} but for ${}^{16}$C beam  energy of
200 (left) and 300 (right) MeV/nucleon and scattering angles
$(\Theta_B=0.0,0.2,0.4^\circ,\varphi_B=0^\circ)$ and
$(\Theta_p=45^\circ,\varphi_p=180^\circ)$. }
\end{figure}

In Fig.~\ref{fig:p2-300} we fix $\Theta_p = 45^\circ$ where the SSA
is expected to be at its best but vary
 $\Theta_B$ and include also results for  $E_\rho^i/16 = 200$ MeV.
At this lower energy the difference between MS-3b and SSA-3b is
more significant, around  15\% at the peak while
at $E_\rho^i/16 = 300$ MeV  it remains below 10\%. 
Thus, these results confirm once more that SSA is more reliable
at higher energies. On the other hand, 
sensitivity to the treatment of $t_{pn}$ energy-dependence is slightly
more pronounced at lower energies:
the difference between SSA-4b and OIA-4b reaches  11\% at
$E_\rho^i/16 = 200$ MeV as compared to 9\% at $E_\rho^i/16 = 300$ MeV.
A further message from Fig.~\ref{fig:p2-300} is that at higher
energy the differential cross section is more sharply peaked
around the QFS point as it decreases faster with increasing $\Theta_B$.

\begin{figure}[!]
\begin{center}
\includegraphics[scale=0.62]{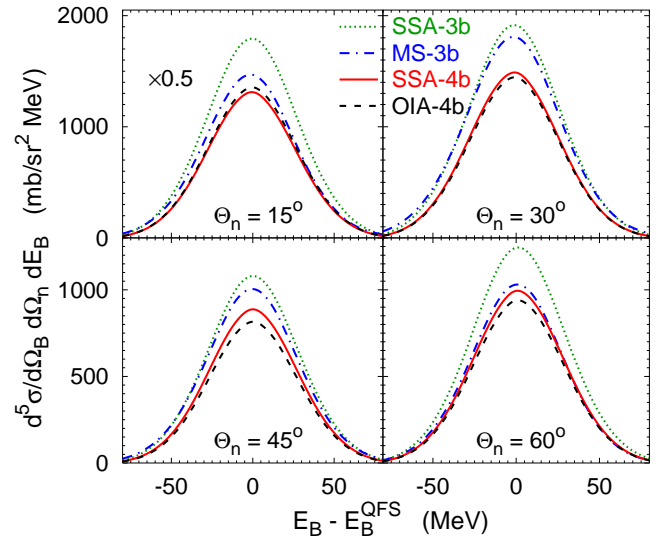}
\end{center} 
\caption{ \label{fig:n300} (Color online) 
Same as Fig.~\ref{fig:p300} but a neutron is detected
instead of a proton  at angles
$(\Theta_n=15,30,45,60^\circ,\varphi_n=180^\circ)$. }
\end{figure}

In Fig.~\ref{fig:n300} we show the results at $E_\rho^i/16 = 300$ MeV
where a neutron is detected instead of a proton. As in Fig.~\ref{fig:p300}
we fix $\Theta_B = 0^\circ$, $\varphi_B= 0^\circ$, $\varphi_n= 180^\circ$, 
 and vary $\Theta_n$. In this case the best agreement between
MS-3b and SSA-3b, about 6\%, is observed at intermediate angles
$\Theta_n = 30^\circ$ and $45^\circ$ while at 
$\Theta_n = 15^\circ$ and $60^\circ$ the difference gets above 20\%.
The effect of OIA is most sizable at $\Theta_n = 45^\circ$
reaching 8\%.

As can be concluded from the above results,
the agreement between MS-3b and SSA-3b, i.e., the reliability of the SSA, 
depends not only on the reaction energy but also on the 
kinematical configuration.
This dependence can be understood by inspecting the $nB$ and $pB$ 
relative energies 
in the final three-cluster breakup state, $E_{nB}$ and  $E_{pB}$.
The values corresponding to QFS peaks in Fig.~\ref{fig:p300} 
are  $E_{nB} = 258$, 207, 137, and 66 MeV and
$E_{pB} = 19$, 70, 141, and 211 MeV for
$\Theta_p=15$, 30, 45, and $60^\circ$, respectively.
The values for Fig.~\ref{fig:n300} are obtained by simply
interchanging $E_{nB}$ and  $E_{pB}$.
Thus, the SSA appears to fail when either $E_{nB}$ or  $E_{pB}$ is not 
large enough, with $E_{pB}$ being more decisive, possibly
 due to $pB$ Coulomb interaction.
A detailed investigation of the MS-3b contributions reveals that,
depending on $E_{pB}$ and  $E_{nB}$,
its difference to SSA-3b is dominated by 2nd order terms 
$t_{pB}G_0t_{pn}$ or $t_{nB}G_0t_{pn}$.

We also note that at the QFS peaks the relative $pn$ energy is
around 150 MeV; thus, it is still in the regime where
the underlying $pn$ potential is well constrained by the experimental
data.

The OIA as given by Eq.~(\ref{eq:t3oi})
results in higher values of the energy $e_{12}$ at which
the $pn$ transition operator $t_{pn}$ is evaluated
as compared to the original SSA (\ref{eq:e3}).
This might lead to smaller  $t_{pn}$ values in OIA since the matrix elements
of $t_{pn}$ in average decrease with increasing energy much like
the total $pn$ cross section does. This may explain why the
OIA-4b in most cases underestimates the  SSA-4b predictions.
Furthermore, we conjecture that the use of the overlap integrals
in many-body reaction models would have qualitatively similar effect,
i.e., would reduce the predicted cross sections.


\section{Four-cluster breakup \label{sec:4c}}

A kinematically complete measurement of the four-cluster breakup
requires the detection of three-clusters at least. As this is extremely
difficult, semi-inclusive observables like momentum distributions are
usually measured. Nevertheless, here we present results
for the fully exclusive differential cross section that serves as a
intermediate step for the calculation of semi-inclusive cross sections.
With this goal in mind it is advantageous to
 choose the relative momenta   as kinematical variables, i.e.,
\begin{subequations} \label{eq:jacobi1}
\begin{align}  
\mathbf{k}_x = {}& \frac12(\mathbf{k}_{n_1} -\mathbf{k}_{n_2}), \\
\mathbf{k}_y = {}& \frac{2m_n\mathbf{k}_p - 
m_p(\mathbf{k}_{n_1}+\mathbf{k}_{n_2})}{m_p+2m_n}, \\
\mathbf{k}_z = {}& \frac{(m_p+2m_n)\mathbf{k}_A -
m_A(\mathbf{k}_p+\mathbf{k}_{n_1}+\mathbf{k}_{n_2})}{M},
\end{align}
\end{subequations}
where the subscripts $n_1$ and $n_2$ distinguish between the two neutrons
and $M=m_A+m_p+2m_n$.
The associated relative energies are
$E_x = k_x^2/2\mu_x$, $E_y = k_y^2/2\mu_y$, and $E_z = k_z^2/2\mu_z$
 with reduced masses $\mu_x = m_n/2$,  $\mu_y = 2m_nm_p/(m_p+2m_n)$,
and $\mu_z = m_A(m_p+2m_n)/M$. For example,
$\mathbf{k}_z$ and $E_z\mu_z/m_A$ are the momentum and energy 
of the nuclear core $A$ in the four-particle center-of mass (c.m.) system, 
while $\mathbf{k}_x$ and $E_x$ are the two-neutron relative
momentum and energy.
Owing to momentum and energy conservation there are
eight independent kinematic variables; we choose them
as the polar and azimuthal scattering angles 
$\Omega_j = (\Theta_j,\varphi_j)$ with $j=x,y,z$ 
and two energies $E_x$ and $E_z$. 
In this representation the spin-averaged eightfold differential cross section
is 
\begin{equation} \label{eq:d8s}
\begin{split}
\frac{d^8\sigma}{d\Omega_x d\Omega_y  d\Omega_z dE_x dE_z} = {} &
 \frac{ (2\pi)^4}{V g_i}
\sum_{m_s} | \langle \Phi_{0} |  T_{0 \rho} | \Phi_{\rho} \rangle|^2 \\
& \times  \mu_x \mu_y \mu_z k_x k_y k_z,
\end{split}
\end{equation}
where, as in Eq.~(\ref{eq:d5s}),
the sum runs over  all initial and final spin states.
The single-particle cross section for the core $A$ in the c.m. frame
can simply be obtained as 
\begin{equation} \label{eq:d3sa}
\frac{d^3\sigma_{\rm c.m.}}{d\Omega_A dE_A} = 
\int d\Omega_x d\Omega_y dE_x \, \frac{m_A}{\mu_z} \,
\frac{d^8\sigma}{d\Omega_x d\Omega_y  d\Omega_z dE_x dE_z}.
\end{equation}

\begin{figure}[!]
\begin{center}
\includegraphics[scale=0.62]{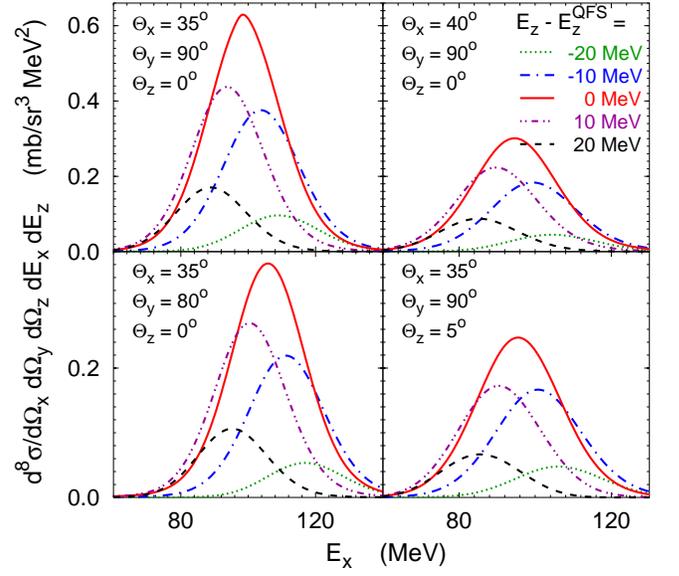}
\end{center} 
\caption{ \label{fig:q300} (Color online) 
Differential cross section for ${}^{16}$C$(p,pnn){}^{14}$C reaction
at  $E_\rho^i/16 = 300$ MeV
as a function of the relative $nn$ energy $E_x$
for selected values of $E_z$ near $pn$ QFS kinematics
and angles $\Theta_x$, $\Theta_y$, and $\Theta_z$ whereas
$\varphi_x=\varphi_y=\varphi_z=0^\circ$.}
\end{figure}

The $pn$ QFS implies $\mathbf{k}_{A} \approx \mathbf{k}_{A}^{\rm QFS}
= (m_A/m_\rho) \mathbf{k}_{\rho}^i$ and 
$\mathbf{k}_{n_2} \approx \mathbf{k}_{n_2}^{\rm QFS}
= (m_n/m_\rho) \mathbf{k}_{\rho}^i$ but also
vanishing relative momentum $ \mathbf{p}_{n_2A} = (m_A\mathbf{k}_{n_2} -
m_n \mathbf{k}_{A})/(m_A+m_n)$ and relative
energy $E_{n_2A}$
(here for simplicity we assume that $n_1$ is knocked out,
but in practical calculations the amplitudes are antisymmetric
with respect to the two neutrons).
 Thus, due to low relative $nA$ energy the SSA is expected to be
less accurate than for the three-cluster breakup in suitable kinematics.
Nevertheless, in Fig.~\ref{fig:q300} we show the results for the
fully exclusive differential cross section (\ref{eq:d8s})
of the  $p+{}^{16}$C$ \to p+n+n+{}^{14}$C reaction
at $E_\rho^i/16 = 300$ MeV. We fix the azimuthal angles
$\varphi_x=\varphi_y=\varphi_z=0^\circ$ and vary the polar angles
near a pronounced QFS peak at 
$\Theta_x=35^\circ$, $\Theta_y=90^\circ$, and $\Theta_z=0^\circ$.
The cross section is shown as a function of the
relative $nn$ energy $E_x$ for selected values of $E_z$ around
$E_z^{\rm QFS}= [m_p^2m_A^2/(\mu_zm_\rho M^2)]E_\rho^i$.
The results demonstrate that the 
differential cross section decreases rapidly whenever kinematical
conditions deviate from QFS.

Finally, we note that the reliability of the SSA predictions for the 
semi-inclusive cross section (\ref{eq:d3sa}) may be even more limited.
The integration in Eq.~(\ref{eq:d3sa}) unavoidably includes
regions of phase space with low relative $pA$ energy where the higher
order terms omitted in the SSA may be significant.

\section{Summary \label{sec:sum}}

Starting with full four-particle AGS equations for the transition 
operators we derived three- and four-cluster breakup
amplitudes in the single-scattering approximation.
Numerical calculations were performed for  ${}^{16}$C breakup
on a proton target with  the ${}^{14}$C core and
two-neutron model for the ${}^{16}$C nucleus.
Breakup reactions at 200 and 300 MeV/nucleon energy
near $pn$ QFS conditions were studied.
In the case of the three-cluster breakup an additional
three-body model, although being physically inadequate, 
allowed to estimate the accuracy of the SSA and thereby
identify the kinematical regimes where the SSA is reliable.
These regimes correspond to higher reaction energies and to
higher relative energies
between the composite cluster and any of the nucleons
and are realized at proton or neutron scattering angles
around $45^\circ$. There the accuracy of the SSA becomes
as good as 6\%. Furthermore, 
we have shown that an additional approximation in the
energy dependence of the $pn$ transition operator
is needed to factorize the SSA of the three-cluster breakup
amplitude into $t_{pn}$ and the 
overlap integral of two- and three-particle bound states.
This approximation usually reduces the cross section,
in some cases even up to 10\%.

No SSA reliability test was possible for the four-cluster breakup
but, given the conclusions drawn in the three-cluster case, 
it is expected to be less accurate. 
An extension of the present calculations to include also the
double-scattering terms, especially those between the core 
and neutron at low relative energy, would be highly desirable.

The obtained three- and four-cluster breakup results 
demonstrate that the differential cross section is sharply peaked
at $pn$ QFS point and decreases rapidly 
whenever kinematical conditions deviate from $pn$ QFS.

The present numerical calculations are limited to $p+{}^{16}$C
reactions. However, the formalism is applicable also to other nuclei 
like ${}^{12}$Be,  ${}^{20}$C, and ${}^{24}$O, and, in the case
of the four-cluster breakup, also to 
 ${}^{6}$He,  ${}^{11}$Li, and  ${}^{22}$C.

\begin{acknowledgments}
The author thanks A.~M.~Moro for discussions and 
the Institute of Theoretical Physics and Astronomy of Vilnius
University  for its hospitality during the completion of this work.
The work was partially supported by the
FCT grant PTDC/FIS/65736/2006.
\end{acknowledgments}


\begin{thebibliography}{10}

\bibitem{baye:09a}
D. Baye, P. Capel, P. Descouvemont, and Y. Suzuki, Phys. Rev. C {\bf 79},
  024607  (2009).

\bibitem{esbensen:95a}
H. Esbensen, G.~F. Bertsch, and C.~A. Bertulani,
Nucl.~Phys. {\bf A581}, 107 (1995).

\bibitem{austern:87}
N. Austern, Y. Iseri, M. Kamimura, M. Kawai, G. Rawitscher, and M. Yahiro,
  Phys. Rep. {\bf 154},  125  (1987).

\bibitem{faddeev:60a}
L.~D. Faddeev, Zh.~Eksp.~Teor.~Fiz. {\bf 39},  1459  (1960) [Sov.~Phys. JETP
  {\bf 12}, 1014 (1961)].

\bibitem{alt:67a}
E.~O. Alt, P. Grassberger, and W. Sandhas, Nucl.~Phys. {\bf B2},  167  (1967).

\bibitem{deltuva:07d}
A. Deltuva, A.~M. Moro, E. Cravo, F.~M. Nunes, and A.~C. Fonseca, Phys.~Rev.~C
  {\bf 76},  064602  (2007).

\bibitem{moro:06a}
A.~M. Moro and F.~M. Nunes, Nucl.~Phys. {\bf A767},  138  (2006).

\bibitem{matsumoto:06a}
T. Matsumoto, T. Egami, K. Ogata, Y. Iseri, M. Kamimura, and M. Yahiro, Phys.
  Rev. C {\bf 73},  051602  (2006).

\bibitem{gallardo:09a}
M. Rodr\'{\i}guez-Gallardo, J.~M. Arias, J. G\'omez-Camacho, A.~M. Moro, I.~J.
  Thompson, and J.~A. Tostevin, Phys. Rev. C {\bf 80},  051601  (2009).

\bibitem{moro:pc}
A.~M. Moro, private communication.

\bibitem{deltuva:12c}
A. Deltuva and A.~C. Fonseca, Phys.~Rev.~C {\bf 86},  011001(R)  (2012).

\bibitem{lazauskas:12a}
R. Lazauskas, Phys. Rev. C {\bf 86},  044002  (2012).

\bibitem{crespo:08a}
R. Crespo, A. Deltuva, E. Cravo, M. Rodriguez-Gallardo, and A.~C. Fonseca,
  Phys.~Rev.~C {\bf 77},  024601  (2008).

\bibitem{grassberger:67}
P. Grassberger and W. Sandhas, Nucl. Phys. {\bf B2},  181  (1967); E. O. Alt,
  P. Grassberger, and W. Sandhas, JINR report No. E4-6688 (1972).

\bibitem{yakubovsky:67}
O.~A. Yakubovsky, Yad. Fiz. {\bf 5},  1312  (1967) [Sov. J. Nucl. Phys. {\bf
  5}, 937 (1967)].

\bibitem{deltuva:07a}
A. Deltuva and A.~C. Fonseca, Phys.~Rev.~C {\bf 75},  014005  (2007).


\bibitem{deltuva:12e}
A. Deltuva, Few-Body Syst.  (2012), DOI:10.1007/s00601-012-0477-0.

\bibitem{deltuva:12a}
A. Deltuva, Phys.~Rev.~A {\bf 85},  012708  (2012).

\bibitem{alt:80a}
E.~O. Alt and W. Sandhas, Phys.~Rev.~C {\bf 21},  1733  (1980).

\bibitem{deltuva:05d}
A. Deltuva, A.~C. Fonseca, and P.~U. Sauer, Phys.~Rev.~C {\bf 72},  054004
  (2005).

\bibitem{deltuva:07b}
A. Deltuva and A.~C. Fonseca, Phys.~Rev.~Lett. {\bf 98},  162502  (2007).

\bibitem{machleidt:01a}
R. Machleidt, Phys.~Rev.~C {\bf 63},  024001  (2001).

\bibitem{deltuva:09d}
A. Deltuva, Phys.~Rev.~C {\bf 79},  054603  (2009).

\bibitem{koning}
A.~J. Koning and J.~P. Delaroche, Nucl. Phys. {\bf A713},  231  (2003).

\bibitem{hammer:07a}
H.~W. Hammer and L. Platter, Eur. Phys. J. A {\bf 32},  113  (2007).

\end{thebibliography}

\end{document}